# EPICS process variables in different subnetworks and different IOCs without the use of the CaGateway

## Rodrigo Bongers

*Abstract*
*This technical document describes the comparison of the EPICS PV gateway and a new solution based on relay of UDP packets using the UDP-HELPER switch feature, iptables and a C program. The solution can be applied on environments that contain multiple sub-networks and a number of IOC's on the same host or multiple IOC's on the same sub-network.*

Experimental Physics and Industrial Control System (EPICS) is a set of software tools and applications that provide a suite for the conception of distributed control systems and the operation of the particle accelerator and beamlines, and this tool can also be used for any purpose that requires automation and control. These distributed control systems normally comprise tens or even hundreds of computers connected in a network to permit communication among them and to provide control and feedback to various parts of the device from a central control room or remotely through the internet.

EPICS uses Client/Server and Publish/Subscribe techniques for communication among various computers. The majority of the servers (called Input/Output Controllers or IOCs) executes I/O tasks both externally as well as locally and publishes a series of variables with information that the system administrator may find relevant to the clients using the Channel Access (CA) network protocol.

Channel Access uses the IP protocol for the publication, search and transfer of data between the various IOCs. There are two current methods for searching and publishing information from other IOCs: the monitored and spot methods. In the first method, the client that requires frequently updated information broadcasts in his subnetwork, searching for the IOC that answers the variable in question. The IOC responds with the IP to which the client should connect to seek the information, and the client is registered in a list of clients interested in the variable who will receive the information whenever it is altered. This system is also used by the "camonitor" command line program. For the spot search/publication, the "caget" and "caput" programs are used; thus, a network broadcast is sent whenever the client needs information or wants to publish a new value for the variable.

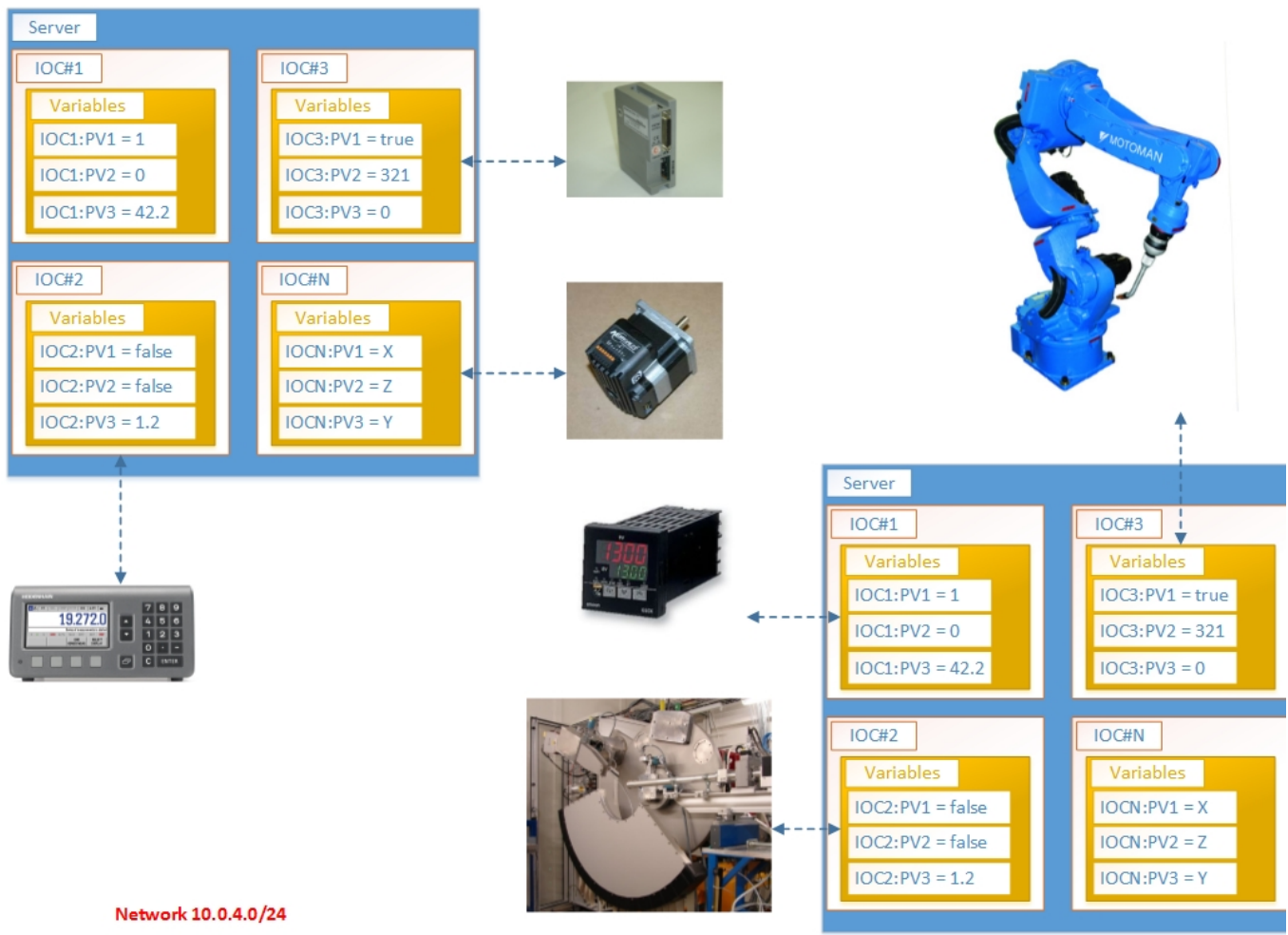

FIGURE 1

When a packet is sent to broadcast, all of the devices that are in the subnetwork in question will receive the packet even if they do not need to use it. Therefore, packet processing is performed for all of those in that subnetwork.

"IP Broadcast: When one or more packets are sent to all hosts or nodes on a network by using a single IP address, it's referred to as IP Broadcast." (Fred Huffman, Practical IP and Telecom for Broadcast Engineering and Operations).

Another important point is that the packets are discarded by those who do not need this information whenever a client sends a request through the "caget" or "caput" programs.

Because all EPICS communication uses broadcasts, the administrator is required to keep the IOCs in the same subnetwork so that all of the servers can consult and act on the variables. However, when a network begins to possess a large volume of equipment and numerous IOCs, the network security is compromised because it is possible to receive and send information to any PV, and the network topology becomes complex. A high broadcast rate can affect other network devices that do not support the same volume of IP datagrams. For EPICS to support operations in multiple subnetworks, an environment variable can be configured in the client so that the packet is directed to specific subnetworks (EPICS_CA_ADDR_LIST="1.2.3.255 8.9.10.255"), or a proxy CA can be configured.

The proxy CA, also known as a Process Variable Gateway, is a program that must be configured in each subnetwork that needs to perform consultations in PVs or must have an interface for each subnetwork, making the installation very complex and requiring investment in hardware and complex software configuration on the client side of the networks. More configurations must be performed for the configured gateways to search the PVs, leaving a large list of IPs to be managed in a complex manner. Figure II demonstrates a configuration that must be performed for the gateway to function.

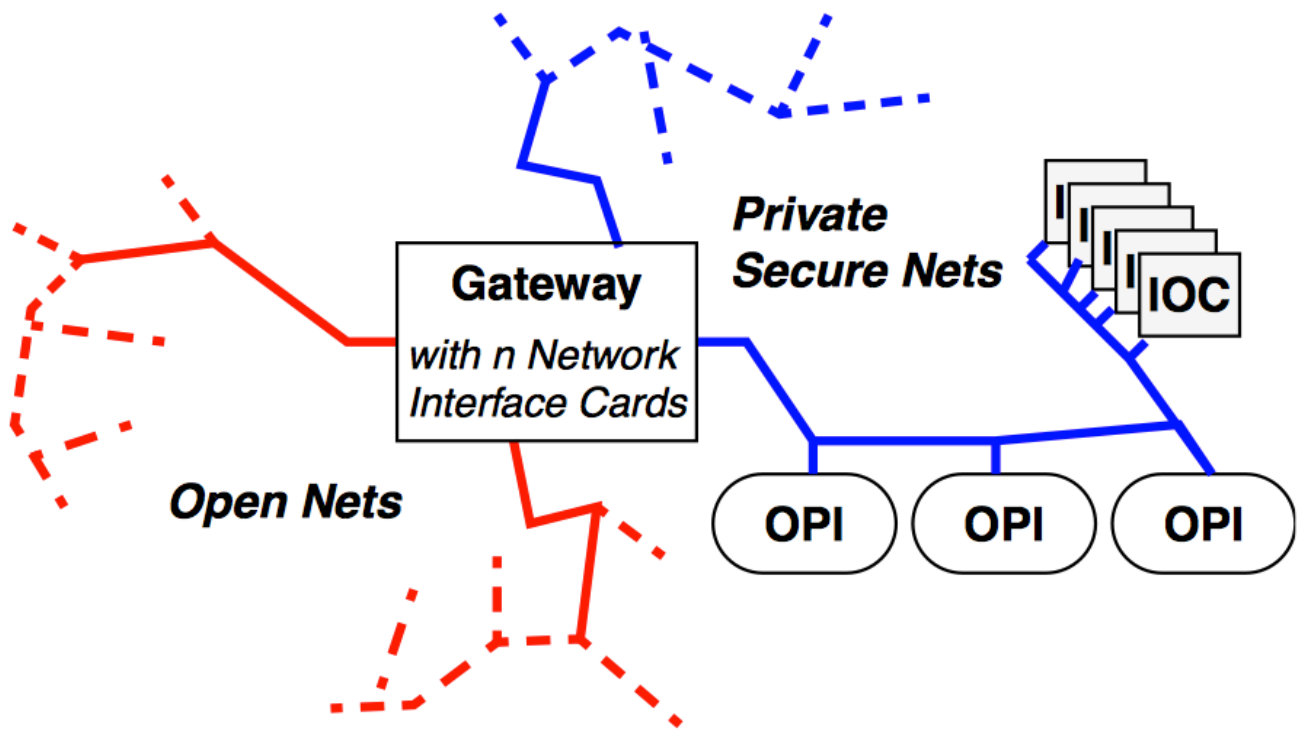

FIGURE 2 - http://www-csr.bessy.de/control/SoftDist/Gateway/download/GatewayUpdate.pdf

Another disadvantage of this technology arises when it is necessary to configure more than one Gateway in more than one network, closing a loop in the topology ( Figure III ).

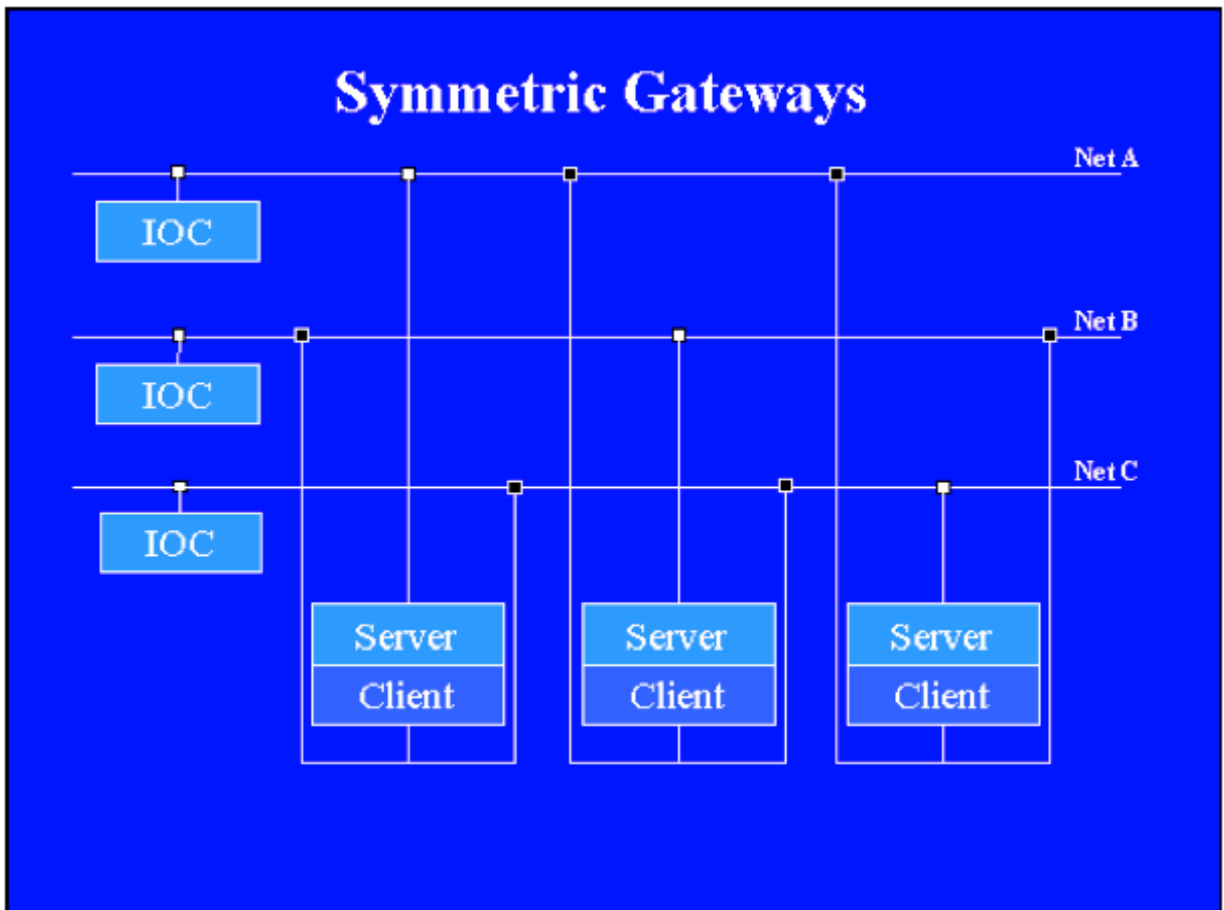

FIGURE 3 -
http://www.aps.anl.gov/epics/EpicsDocumentation/ExtensionsManuals/Gateway/ICALEPCS2005.Gateway.evans.pdf

A PXI system standard (National Instruments Equipment that combines processing and specialized and high performance data acquisition plates) was adopted in the beamlines of the LNLS. The difference is Hyppie, a virtualization system for the use of Labview RealTime and a Linux system in the same hardware. The equipment also supports a large quantity of communication buses and signals, such as Serial RS232 and RS422/485, EtherCat, Ethernet, FPGA and TTL, such that all of the PXIs have EPICS and numerous IOCs configured. Like any successful distributed system, the increase in equipment and consequently in broadcast and human errors in writing, reading and PVs with duplicate names makes the system complex and difficult to troubleshoot.

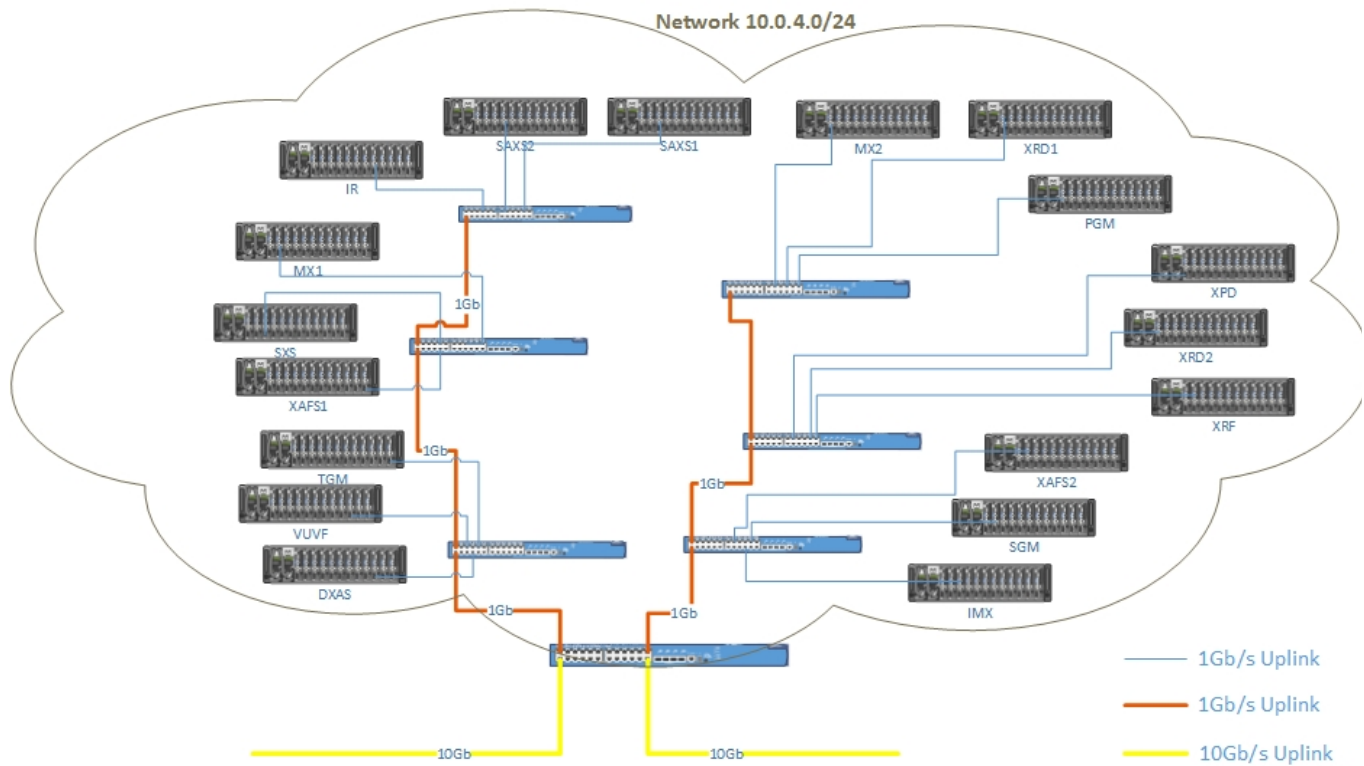

Figure 4 – Network topology with all devices on the same subnetwork

Each beamline has a series of equipment that are controlled via EPICS. The equipment must be protected from human errors because a synchrotron light source emits radiation and the motors are sufficiently strong in some cases to break equipment, such as the sampling port, if abrupt or incorrect movements are made. With the system operating in the same subnetwork, accidents could occur, including the movement of motors from another beamline, the opening of the shutter, the opening of the system of slots and the loss of research that was occurring at the time, which can normally last weeks.

Together with these difficulties, the demand for the allocation of space in unified storage, band pass for the new detectors and IP numbers, the network modernization project arose to fix the current problems and include a link resilience protocol between the beamline switches. The project considered the following items:
- 7 Core Network Switches
- 18 Local Network Switches for the beamlines
- OSPF Protocol for link resilience
- LACP Protocol for 4 GB/s of the network per beamline
- segmentation of the network departing from 10.0.4.0/24 (255 ips) to 72 networks (18360 ips) with 4 VLANs for each beamline to be distributed in the most convenient manner to the beamline engineers and physicists.
- GVRP/MVRP Protocol for the dynamic addition and removal of ports in the VLANs for support groups that need to be in the same VLAN of the line to receive maintenance.

Thus, the broadcast is limited only within the beamlines, and the more sensitive equipment (for example, the "Powerpack", sets of stepper motor controllers that are more susceptible to broadcasts of large packets) can be in a separate VLAN from the PXI that has EPICS in production. The powerpack is an appliance with Phytron drivers and a Galil controller used for moving the slot motors, monochromators and sampling ports on the beamlines, with a network plate to receive the movement commands and feedback. However, some versions cannot receive Jumbo Frame packets because they do not have the power to process them.

Following project implementation, all of the beamline commands were restricted to the subnetwork line to eliminate the problem of sending and receiving commands from other users from other lines. There is, however, the SOL group (Line operation software), which responsible for the creation and maintenance of IOCs and the administration of the PXIs. The SOL group was in a subnetwork of the group and ceased to visualize all of the beamline variables. Figure V demonstrates how the group is directly connected to the ANE-SUP-S29-2 switch in a 10.2.101.0/24 network. IMX is an example of a segmented beamline, which corresponds to the ANE-INF-IMX switch and to the PXI equipment named IMX1-HOST1 running EPICS.

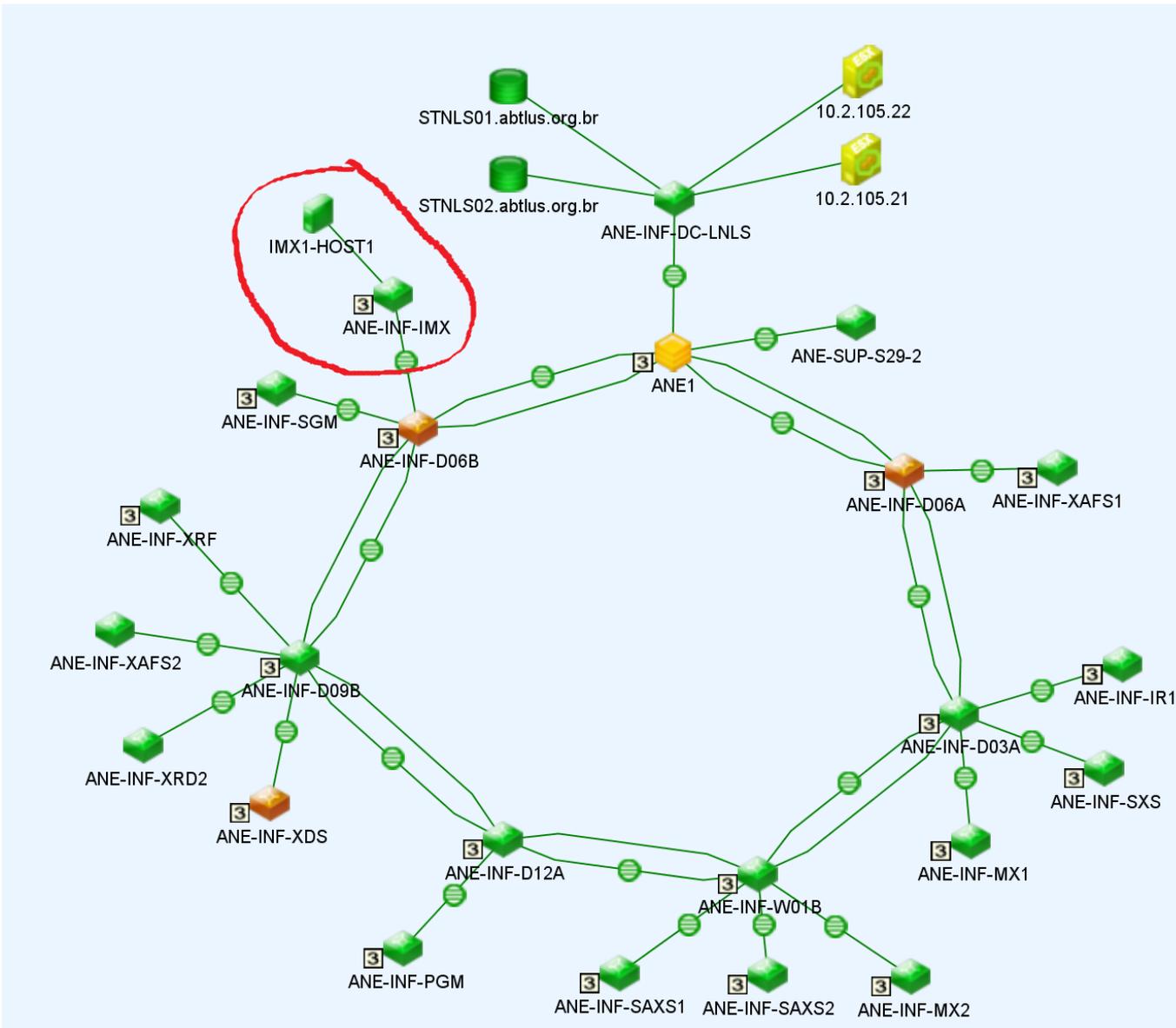

FIGURE 5

The DHCP Protocol uses the same idea to search for a server in the network to acquire an IP; a UDP broadcast is sent to port 67, and the DHCP server responds with an available IP. However, when the server is in a different subnetwork, the RELAY technique is used for the switch to redirect the packet to the server. The technique was replicated in the switch of the SOL group for EPICS packets, enabling the UDP-HELPER in port 5064 and directing to the PXIs from all of the beamlines. The packet relayed by the switch is converted from broadcast to unicast and sent to the server [1].

"UDP Helper (UDPH) functions as a relay agent that converts UDP broadcast packets into unicast packets and forwards them to a specified server. With the UDP Helper function enabled, the device decides whether to forward a received UDP broadcast packet according to the UDP port number of the packet. If the packet needs to be forwarded, the device modifies the destination IP address in the IP header and then sends the packet to the specified destination server. Otherwise, the device sends the packet to its upper layer module."

For each IOC running on the server, a UDP connection is established with the broadcast address of the subnetwork as shown below:

```
[root@IMX1-HOST1 ~]# netstat -anpo |grep 10.2.1.255
udp    0    0 10.2.1.31:34192   10.2.1.255:5065 ESTABLISHED 2798/HostUptime off
(0.00/0/0)
udp    0    0 10.2.1.31:49939   10.2.1.255:5065 ESTABLISHED 22962/galilTest off
(0.00/0/0)
udp    0    0 10.2.1.31:58266   10.2.1.255:5065 ESTABLISHED 2810/PFCU       off
(0.00/0/0)
udp    0    0 10.2.1.31:40006   10.2.1.255:5065 ESTABLISHED 2808/galilTest  off
(0.00/0/0)
udp    0    0 10.2.1.31:49775   10.2.1.255:5065 ESTABLISHED 2804/galilTest  off
(0.00/0/0)
udp    0    0 10.2.1.31:55678   10.2.1.255:5065 ESTABLISHED 2803/galilTest  off
(0.00/0/0)
udp    0    0 10.2.1.31:50942   10.2.1.255:5065 ESTABLISHED 2799/digital    off
(0.00/0/0)
```

As a consequence, with EPICS, when a packet is converted from Broadcast to Unicast, the PXI cannot additionally process the packet for all of the IOCs that are running on the machine, only for the last IOC that established a connection with the broadcast through the kernel security measure. The section below presents the communication from the last IOC that is running on the machine with the PV named IMX1-HOST1:

The command in the client was as follows:

```
[root@TesteRHEpics linux-x86_64]# ./caget IMX1-HOST1
IMX1-HOST1                    0.0191667
```

[root@IMX1-HOST1 HostUptime]# tcpdump -i br0 -n port 5064
tcpdump: verbose output suppressed, use -v or -vv for full protocol decode
listening on br0, link-type EN10MB (Ethernet), capture size 96 bytes

```
10:21:14.615980 IP 10.2.105.156.46702 > 10.2.1.31.ca-1: UDP, length 64
10:21:14.952239 IP 10.2.105.171.35687 > 10.2.1.31.ca-1: UDP, length 48
10:21:14.952337 IP 10.2.1.31.ca-1 > 10.2.105.171.35687: UDP, length 40
10:21:14.953908 IP 10.2.105.171.35948 > 10.2.1.31.ca-1: S 4078395848:4078395848(0)
win 14600 <mss 1460,sackOK,timestamp 1819293380 0,nop,wscale 7>
10:21:14.953934 IP 10.2.1.31.ca-1 > 10.2.105.171.35948: S 2575747268:2575747268(0)
ack 4078395849 win 5792 <mss 1460,sackOK,timestamp 6032359 1819293380,nop,wscale 7>
10:21:14.954130 IP 10.2.105.171.35948 > 10.2.1.31.ca-1: . ack 1 win 115
<nop,nop,timestamp 1819293380 6032359>
10:21:14.954300 IP 10.2.105.171.35948 > 10.2.1.31.ca-1: P 1:113(112) ack 1 win 115
<nop,nop,timestamp 1819293380 6032359>
10:21:14.954315 IP 10.2.1.31.ca-1 > 10.2.105.171.35948: . ack 113 win 46
<nop,nop,timestamp 6032360 1819293380>
10:21:14.954636 IP 10.2.1.31.ca-1 > 10.2.105.171.35948: P 1:33(32) ack 113 win 46
<nop,nop,timestamp 6032360 1819293380>
10:21:14.954833 IP 10.2.105.171.35948 > 10.2.1.31.ca-1: . ack 33 win 115
<nop,nop,timestamp 1819293381 6032360>
10:21:14.955135 IP 10.2.105.171.35948 > 10.2.1.31.ca-1: P 113:129(16) ack 33 win
115 <nop,nop,timestamp 1819293381 6032360>
10:21:14.955254 IP 10.2.1.31.ca-1 > 10.2.105.171.35948: P 33:73(40) ack 129 win 46
<nop,nop,timestamp 6032361 1819293381>
10:21:14.955813 IP 10.2.105.171.35948 > 10.2.1.31.ca-1: P 129:145(16) ack 73 win
115 <nop,nop,timestamp 1819293382 6032361>
10:21:14.955819 IP 10.2.105.171.35948 > 10.2.1.31.ca-1: F 145:145(0) ack 73 win 115
<nop,nop,timestamp 1819293382 6032361>
10:21:14.955918 IP 10.2.1.31.ca-1 > 10.2.105.171.35948: P 73:89(16) ack 146 win 46
<nop,nop,timestamp 6032361 1819293382>
10:21:14.956024 IP 10.2.1.31.ca-1 > 10.2.105.171.35948: F 89:89(0) ack 146 win 46
<nop,nop,timestamp 6032361 1819293382>
```

```
10:21:14.956224 IP 10.2.105.171.35948 > 10.2.1.31.ca-1: . ack 90 win 115
<nop,nop,timestamp 1819293382 6032361>
```

In the section below, an IOC was the first to be initiated and to which the client did not receive a response from the PV

The command in the client was as follows:

```
[root@TesteRHEpics linux-x86_64]# ./caget IMX:DMC4:m1
Channel connect timed out: 'IMX:DMC4:m1' not found.
```

```
10:21:18.724632 IP 10.2.105.171.44256 > 10.2.1.31.ca-1: UDP, length 48
10:21:18.756589 IP 10.2.105.171.44256 > 10.2.1.31.ca-1: UDP, length 48
10:21:18.820475 IP 10.2.105.171.44256 > 10.2.1.31.ca-1: UDP, length 48
10:21:18.948576 IP 10.2.105.171.44256 > 10.2.1.31.ca-1: UDP, length 48
10:21:19.204807 IP 10.2.105.171.44256 > 10.2.1.31.ca-1: UDP, length 48
10:21:22.244620 IP 10.2.101.47.37520 > 10.2.1.31.ca-1: UDP, length 112
```

Because the solution was not functioning according to expectations and did not meet the prerequisites, another manner of fixing the problem was applied, again transforming the packet into broadcast soon after it arrived in the Linux routing table before being processed by the system in the pre-routing table. This table processes all of the packets that meet the requirements of the created rule and performs the alteration desired by the administrator. For this, the following rule was created:

iptables -t nat -A PREROUTING -p udp ! -s `ifconfig br0 |grep "inet addr" |awk '{print $2}' |cut -f2 -d: |cut -f1-3 -d.`.0/24 --dport 5064 -j DNAT --to <u>255.255.255.255:5064</u>

BROADCAST--->UDP-HELPER---UNICAST--->PREROUTING----BROADCAST--->PXI

Thus, the clients began to receive PVs from all of the IOCs that exist within the server, as shown in the commands below:

```
[root@TesteRHEpics linux-x86_64]# ./caget IMX:DMC4:m1
IMX:DMC4:m1                    -2.06e-05
[root@TesteRHEpics linux-x86_64]# ./caget IMX1-HOST1
IMX1-HOST1             0.122222
[root@TesteRHEpics linux-x86_64]# ./caget IMX:DMC4:m2
IMX:DMC4:m2                    -1.47e-05
[root@TesteRHEpics linux-x86_64]# ./caget IMX:DMC4:m3
IMX:DMC4:m3             0.002496
```

In fact, the IPTABLES rule altered the header with the destination IP in the pre-routing table, such that all of the IOCs that ran on this server accepted the packet. However, packet alteration did not redirect to broadcast; it only altered the packet such that the machine that contained the rule could accept it. This change resolved the problem on the beamline with only one server and various IOCs but did not solve the problem for beamlines where there were two or more servers with EPICS and other PVs to be consulted in the IOCs. For redirecting to actually occur, it would be necessary to receive the RAW Packet, recreate the IP header and redirect the packet to the output interface of the network, maintaining the origin address so that the IOC responded to whoever actually consulted the PV.

Two outputs were found to solve the problem. The first is a program called "socat", which can be installed on any Linux distribution and manipulates sockets in various manners, making it possible to receive commands in STDIN, listen on pre-defined ports, broadcast, multicast, TCP, UDP, RAW, IPv6 and send the output to any previously cited option as well as to serial ports, with the option to perform a fork to receive more than one request. The command below was used for the successful tests:

```
socat UDP4-LISTEN:6064,fork,range=10.2.105.0/24 UDP4-DATAGRAM:255.255.255.255:5064,broadcast
```

The above command listens to packets from port 6064 (from the IPTABLES that receives packets from 5064 and directs them to 6064 if the packet is not from the same subnetwork), creates a sub-process if the packet is from the SOL group subnetwork and sends the UDP datagram again to broadcast in port 5064, maintaining the origin IP. With this, the segmentation from the network and the UDP-HELPER begin to function as observed in the command below:

```
[root@TesteRHEpics linux-x86_64]# ./caget IMX:DMC4:m1
IMX:DMC4:m1                    -2.06e-05

[root@TesteRHEpics linux-x86_64]# ./caget IMX1-HOST1
IMX1-HOST1                     155.836

[root@TesteRHEpics linux-x86_64]# ./caget IMX:DMC4:m3
IMX:DMC4:m3                    0.002496
```

The above PVs are on separate servers in the same 10.2.1.0/24 subnetwork. The computer that performed the query (TesteRHEpics) has the IP address 10.2.105.171, and the switch to which it is connected is performing UDP-HELPER for the IP 10.2.1.31, which is where the IOCs, IPTABLES with the pre-routing rule and "socat" are being executed.

The response time of the PVs is another necessity for some beamlines. The times are below 0.075 s for consultation in the same subnetwork or on the server in which the IOC is being executed. With socat and iptables running, the times are much longer than expected because a sub-process must be created for each consultation, overloading the server executing the task and slowing the responses.

To improve times, it is necessary to develop a low level programming language that meets the performance requirements, has low resource consumption for activities with the socket and is totally compatible with Linux. C is the ideal language for this scenario. A program that reuses the RAW packet and finally recreates it with the IP of the original origin and the broadcast destination was developed.

Figure VI presents the times for the three scenarios; the consultation time in the same subnetwork is the ideal comparison time for the other two solutions. The times were measured more than once, and a measurement was calculated for the final result.

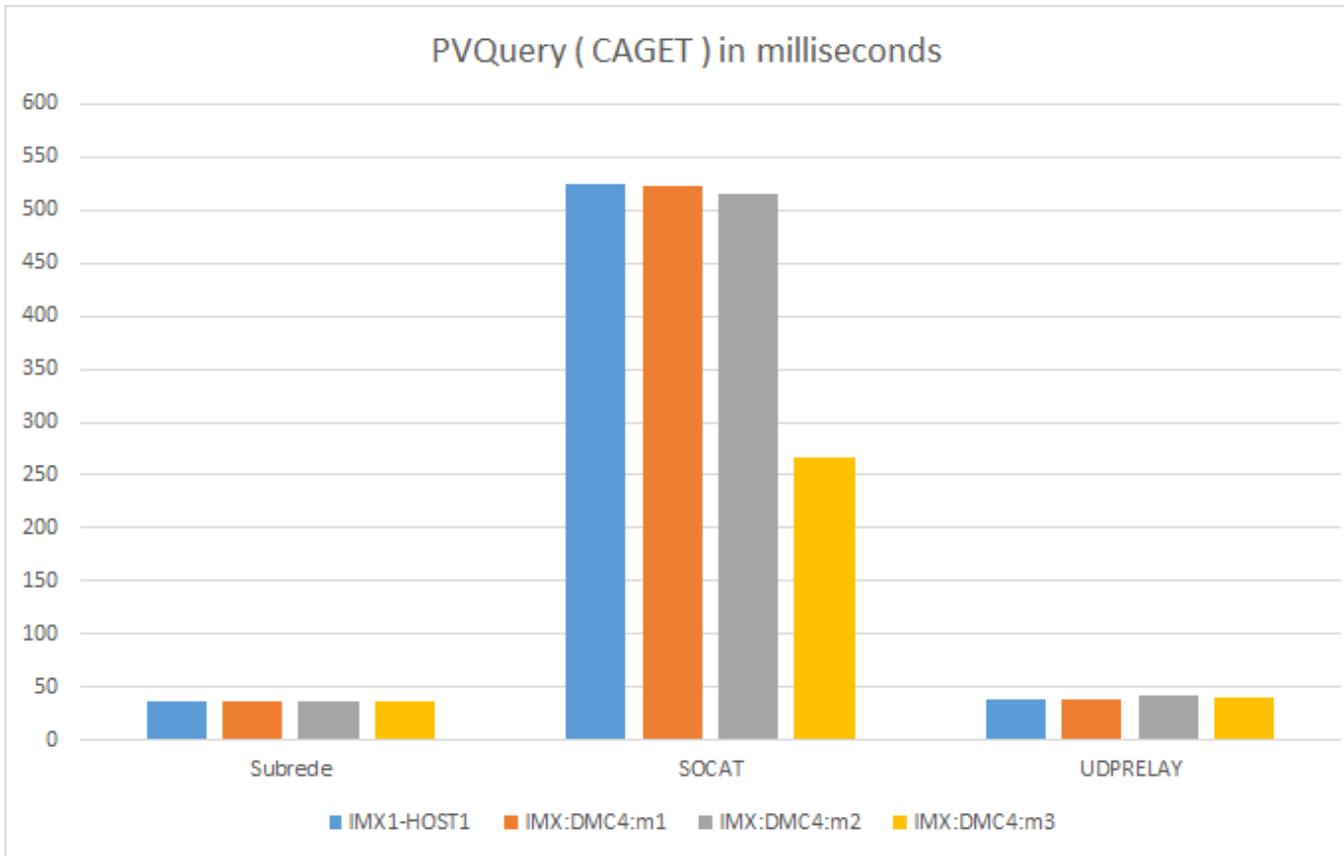

FIGURE 6

If security measures must be implemented in the PV consultations to create personalized accesses, this can be done directly in the IOC or in the VLAN on which the EPICS server is installed so that the adopted solution is at the same level as any other solution adopted for the EPICS platform.

Conclusion

The use of UDPRELAY solution keeping all the requirements of security, low latency and high performance on implementation of a distributed system that meets the current needs of synchrotron beamlines system. With the available computational power and analysis of packets carried by a native linux firewall, the solution does not impact server performance. Furthermore, this system allows the network administrator to act without creating difficulties for engineers or analysts who develop and maintain the EPICS tool.